\documentclass[%
 aip,
 amsmath,amssymb,
 reprint,%
]{revtex4-1}

\usepackage{graphicx}
\usepackage{dcolumn}
\usepackage{bm}

\usepackage[utf8]{inputenc}
\usepackage[T1]{fontenc}
\usepackage{mathptmx}
\usepackage{etoolbox}
\usepackage[table]{xcolor}
\usepackage{booktabs}
\usepackage{siunitx}
\usepackage{tcolorbox}
\usepackage{nicematrix,tikz}

\newcommand{\ebg}[1]{\cellcolor{blue!20}{#1}}
\newcommand{\nbg}[1]{\cellcolor{red!20}{#1}}
\newcommand{\enbg}[1]{\cellcolor{green!20}{#1}}

\makeatletter
\def\@email#1#2{%
 \endgroup
 \patchcmd{\titleblock@produce}
  {\frontmatter@RRAPformat}
  {\frontmatter@RRAPformat{\produce@RRAP{*#1\href{mailto:#2}{#2}}}\frontmatter@RRAPformat}
  {}{}
}%
\makeatother
\begin{document}

\preprint{AIP/123-QED}

\title{$^{13}$C and $^{19}$F Nucleus-Electron Correlation and Self-Energies}

\author{Janina Vohdin}
\affiliation{Institute of Quantum Materials and Technologies, Karlsruhe
Institute of Technology (KIT), Kaiserstra\ss{}e 12, 76131 Karlsruhe,
Germany}

\author{Christof Holzer}
\affiliation{Institute of Quantum Materials and Technologies, Karlsruhe
Institute of Technology (KIT), Kaiserstra\ss{}e 12, 76131 Karlsruhe,
Germany}
\email[Email for correspondence: ]{holzer@kit.edu}

\date{\today}

\begin{abstract}
We present a theoretical and numerical study of the correlation between 
electrons and the fermionic $^{13}$C and $^{19}$F nuclei. 
We use the random-phase approximation (RPA) as a valuable tool in obtaining 
these correlation energies. A special connection between the RPA and 
second-order perturbation theory for the inter-fermionic interaction
is outlined. Subsequently, Green's function based $GW$ self-energies 
are evaluated for the nuclear densities. The strong influence of 
self-interaction errors is outlined, and vertex corrections are shown 
to be strictly necessary to obtain reasonable results. The theoretical and technical 
requirements for a quantum mechanical treatment of $^{13}$C and $^{19}$F nuclei 
are also addressed in this work, thereby facilitating further research in this area.
\end{abstract}

\maketitle

\section{Introduction}
\label{sec:intro}

Since the introduction of the Born--Oppenheimer (BO) approximation, 
molecular quantum mechanics strongly relies on it.\cite{Born.Oppenheimer:Zur.1927} 
Born and Oppenheimer have tediously proven this to be an excellent
approximation under most circumstances. Confirmed by countless investigations and 
real-life proof-of-work of the BO approximation, 
there is no need to abandon it for the majority of quantum chemical applications. 
Our curiosity nevertheless leads to many questions, 
such as how the quantum chemical world functions beyond the BO approximation.
\cite{Tubman.Kylanpaa.ea:Beyond.2014,Li.Requist.ea:Density.2018,
Kolesov.Kaxiras.ea:Density.2018}
For example, understanding the behavior of two fermions in the high-density limit is 
crucial for understanding the correct asymptotic behavior of the 
correlation energy functionals.\cite{Holzer.Franzke:General.2025}
Although so far two-electron systems
consisting of one spin-up and one spin-down electron have been used
to parameterize the high-density limit of correlation,
\cite{Perdew.Ruzsinszky.ea:Gedanken.2014,Perdew.Burke.ea:Generalized.1996,
Tao.Perdew.ea:Climbing.2003,
Umrigar.Gonze:Accurate.1994,Holzer.Franzke:General.2025} additional information on the 
behavior of the correlation between two different fermions 
is valuable. Investigations of the electron-proton correlation
\cite{Matsushita:Model.1995,Pak.Hammes-Schiffer:Electron-Proton.2004,
Swalina.Pak.ea:Alternative.2005,
Swalina.Pak.ea:Explicit.2006,Sirjoosingh.Pak.ea:Multicomponent.2012,
Brorsen.Schneider.ea:Alternative.2018,Tao.Yang.ea:Multicomponent.2019,
Holzer.Franzke:Beyond.2024,Chen.Yang:Nucleus-electron.2021} 
are more widely available, though no equivalent investigations exist 
for heavier fermions. In this work, we will therefore discuss important 
points about the interactions of fermionic $^{13}$C and $^{19}$F nuclei 
with their surrounding electrons.
We therefore seek to present an initial implementation 
of a general density functional theory framework applicable to heavier 
fermionic nuclei. We then focus on the principal correlation between electrons 
and the $^{13}$C and $^{19}$F nuclei, as well as the binding energies 
of the nuclei to the molecular systems. Despite the explicit nucleus-electron correlation
being considered to be insignificant for standard quantum chemical applications, 
recent applications of generalized fermion density functional
theory have highlighted the importance of understanding these kinds
of interactions for the design of density functional approximations.
\cite{Holzer.Franzke:General.2025}
Our interest in the details of electron–nucleus correlation has been 
further stimulated by recent observations of quantum tunneling involving 
heavier nuclei.\cite{Liao.Bowers.ea:Fluorination.2015,
Nandi.Sucher.ea:Ping-Pong.2018,Nunes.Eckhardt.ea:Competitive.2019,
Muller.Bader.ea:Experimental.2025} For the latter, explicit simulations
of quantum nuclei embedded in electronic simulations will be of 
importance, but still require to solve a multitude of theoretical and
technical issues. While the latter remains a goal for future work, 
this study aims to establish a solid and systematically improvable 
foundation toward its realization, providing the theoretical framework 
and methodological groundwork necessary for subsequent developments.

\section{Theory}
\label{sec:theory}

Multicomponent DFT has been described in detail in the literature,
\cite{PhysRevA.84.052113, Brorsen.Yang.ea:Multicomponent.2017,
Yu.Hammes-Schiffer:Nuclear-Electronic.2020, Pavosevic.Culpitt.ea:Multicomponent.2020, 
HammesSchiffer:Nuclearelectronic.2021,Goli.Shahbazian:Two-component.2022}
and an application of the method to $^{13}$C and $^{19}$F nuclei
is straightforward, simply requiring one to modify the 
corresponding kinetic and potential energy terms by the corresponding
mass and charge of the used particle. This is also true for 
Coulomb integrals, which need to be adapted to charges larger than 
$\pm 1$, which we consider to be trivial. In  this work, we further use
Gaussian basis functions, indicated by the Greek letter $\chi$, while
for one-particle Kohn--Sham (KS) states the Latin letter $\phi$ is used.
Thus, the KS states of two fermions in the Gaussian basis set approximation read
\begin{align}
q' = \phi_q'(x) = & \sum_{\mu }c_{\mu \sigma,q}' \chi_{\mu}'(r) \\
q'' = \phi_q''(x) = & \sum_{\mu}c_{\mu \sigma,q}'' \chi_{\mu}''(r)\text{.}
\end{align}
As usual, $\sigma$ denotes the spin of the fermion and $x$ is a combined
spin-spatial coordinate. Unprimed indices $p$ will be used to denote combined 
spaces of both fermions, 
\begin{equation}
    q = \left\{ q', q'' \right\} = \left\{ \phi_q'(x), \phi_q''(x) \right\} 
\end{equation}
while separate spaces will be primed, as indicated by $q'$ and $q''$. 
The sets of one-electron functions $q'$ and $q''$ are obtained from 
multicomponent KS-DFT as outlined in literature.
\cite{PhysRevLett.86.2984, PhysRevLett.101.153001,
PhysRevA.78.022501}

\subsection{Correlation Energies Between Fermions}
\label{subsec:N6corr}

To access the correlation energy between two different fermions, 
as for example electrons and $^{13}$C or $^{19}$F nuclei, 
we use the random-phase approximation (RPA) as a valuable tool.
As outlined in a recent review, \cite{Chen.Voora.ea:Random.2017}
the adiabatic connection (AC) is a key tool in obtaining correlation energies.
Within the AC, a single coupling strength parameter, denoted $\alpha$, 
is used to switch continuously from the non-interacting KS system 
to the many-fermion system. In the multicomponent framework, 
the coupling strength dependent AC Hamiltonian reads
\begin{equation}
\begin{split}
    \hat{H}_{\alpha}(\rho_e, \rho_f) & = \hat{T}_e + \hat{T}_f + \hat{V}_{ne} + \hat{V}_{nf} \\
    & + \hat{V}_{\alpha}(\rho_e, \rho_f) + \alpha (\hat{V}_{ee} + \hat{V}_{ef} + \hat{V}_{ff} ) \text{,}
\end{split}
\end{equation}
where $\hat{T}$ is the kinetic energy operator of the electrons and nuclei respectively, $\hat{V}_{ne}$ and 
$\hat{V}_{nf}$ are the nucleus-electron and nucleus-fermion Coulomb operators, 
$\hat{V}_{\alpha}(\rho_e, \rho_f)$ is a one-fermion local operator determined 
by the density constraint, and $\alpha (\hat{V}_{ee} + \hat{V}_{ef} + \hat{V}_{ff})$
collects all fermion-fermion Coulomb interactions.
Similar to standard Born-Oppenheimer KS and AC theory, 
$\hat{V}_{\alpha}(\rho_e, \rho_f)$ turns into the Hartree, exchange, and correlation
potential in the non-interacting limit.\cite{Chen.Voora.ea:Random-Phase.2017}
This essentially yields the KS determinant as ground-state wavefunction, and
the missing correlation is to be determined as
\begin{equation}
    E = \langle \phi | \hat{H} | \phi \rangle + E^{\text{C}} \text{.}
\end{equation}
Expressing the correlation energy is now straightforward following
the known procedures, obtaining
\begin{equation}
\begin{split}
    E^{\text{C}}(\rho_e, \rho_f) =  \int_0^1 d\alpha 
    & \left( \langle \psi_\alpha (\rho_e, \rho_f) | \hat{V}_{ee} + \hat{V}_{ef} + \hat{V}_{ff} | \psi_\alpha (\rho_e, \rho_f) \rangle \right. \\
    - & \left.  \langle \phi (\rho_e, \rho_f) | \hat{V}_{ee} + \hat{V}_{ef} + \hat{V}_{ff} | \phi (\rho_e, \rho_f) 
    \right) \text{.}
\end{split}
\end{equation}
Using the fluctuation-dissipation theorem, we can re-express
the correlation energy as an imaginary integral, yielding
\cite{Furche:Molecular.2001}
\begin{equation}
\label{eq:corr_chi}
    E^{\text{C}}(\rho_e, \rho_f) = \frac{1}{4 \pi} \int_0^1 d\alpha \, \text{Im} 
    \int_{-\infty}^{\infty} d\omega \frac{\chi_{\alpha}(x_1, x_2, \omega) - \chi_0(x_1, x_2, \omega)}{|\mathbf{r}-\mathbf{r}'|} \text{,}
\end{equation}
with $\chi_{\alpha}(x_1, x_2, \omega)$ being the frequency-dependent
linear response function. Next, we define the multicomponent time-dependent 
density-matrix -- density-matrix response function
\begin{equation}
\label{eq:tdks}
    \boldsymbol{\Pi}_{\alpha}(\omega) = \left[ 
    \begin{pmatrix}
\mathbf{A}' & \mathbf{B}' & \mathbf{C}^{T} & \mathbf{C}^{T} \\
\mathbf{B}' & \mathbf{A}' & \mathbf{C}^{T} & \mathbf{C}^{T} \\
\mathbf{C}  & \mathbf{C}  & \mathbf{A}''   & \mathbf{B}''   \\
\mathbf{C}  & \mathbf{C}  & \mathbf{B}''   & \mathbf{A}'' 
\end{pmatrix}
- (\omega + i\eta)
\begin{pmatrix}
    \mathbf{1} &  \phantom{-}\mathbf{0} & \phantom{-}\mathbf{0} &  \phantom{-}\mathbf{0} \\
    \mathbf{0} & -1 & \phantom{-}\mathbf{0} &  \phantom{-}\mathbf{0} \\
    \mathbf{0} &  \phantom{-}\mathbf{0} & \phantom{-}\mathbf{1} & \phantom{-}\mathbf{0} \\
    \mathbf{0} &  \phantom{-}\mathbf{0} & \phantom{-}\mathbf{0} & -\mathbf{1}
\end{pmatrix}
\right]^{-1} 
\end{equation}
with the matrix elements
\begin{align}
\label{eq:pi_resp}
    A'_{i'a', j'b'} = & (\epsilon_{a'} - \epsilon_{i'}) \delta_{i'j'} \delta_{a'b'} + \alpha (i'a'|b'j') + f^{\text{XC}, \alpha}_{i'a', j'b'}(\omega) \text{,} \\
    A''_{i''a'', j''b''} = & (\epsilon_{a''} - \epsilon_{i''}) \delta_{i''j''} \delta_{a''b''} + \alpha (i''a''|b''j'') + f^{\text{XC}, \alpha}_{i''a'', j''b''}(\omega) \text{,} \\
    B' = & (i'a'|j'b') + f^{\text{XC}, \alpha}_{i'a', j'b'}(\omega) \text{,} \\
    B'' = & (i''a''|j''b'') + f^{\text{XC}, \alpha}_{i''a'', j''b''}(\omega) \text{,} \\
    \label{eq:pi_resp5}
    C = & (i'a'|j''b'') \text{.}
\end{align}
$\chi_{\alpha}(x_1, x_2, \omega)$ can then be obtained 
from equation~\ref{eq:tdks} as
\begin{equation}
    \chi_{\alpha}(\omega, x_1, x_2) = \Pi_{\alpha}(\omega, x_1, x_1, x_2, x_2) \text{,}
\end{equation}
connecting both quantities. While this relation is convenient to 
outline the connection, we prefer to re-express equation~\ref{eq:corr_chi} using 
the Bethe-Salpeter equation, re-formulating equation~\ref{eq:pi_resp} as
\begin{equation}
\label{eq:pi_alpha}
    \boldsymbol{\Pi}_{\alpha}(\omega) = \boldsymbol{\Pi}_{0}(\omega) + 
    \boldsymbol{\Pi}_{0}(\omega) \left( \alpha \mathbf{V} + \mathbf{F}_{\alpha}^{\text{XC}}(\omega)  \right) \boldsymbol{\Pi}_{\alpha}(\omega) \text{.}
\end{equation}
In equation~\ref{eq:pi_alpha}, 
$\boldsymbol{\Pi}_{0}(\omega) = \boldsymbol{\Pi}_{\alpha=0}(\omega)$, 
and $\mathbf{V}$ collects all Coulomb contributions from 
equations~\ref{eq:pi_resp} to \ref{eq:pi_resp5}, while
$\mathbf{F}_{\alpha}^{\text{XC}}(\omega)$ collects the remaining
exchange-correlation contributions from these equations.
This finally leads to the correlation energy being obtained as
\cite{Holzer.Franzke:Beyond.2024,PhysRevB.15.2884,Ren.Rinke.ea:Random-phase.2012,Holzer:Practical.2023}
\begin{equation}
\label{eq:corr_pi}
    E^{\text{C}} = \frac{1}{4\pi} \int_0^1 d\alpha \, \text{Im} \int_{-\infty}^{\infty} \text{Tr} \left[ \mathbf{V} \left( \boldsymbol{\Pi}_{\alpha} (\omega) - \boldsymbol{\Pi}_0(\omega) \right) \right] \text{,}
\end{equation}
using a double integral over imaginary frequencies 
and the coupling strength $\alpha$. As we do not know
about the exact nature of $f^{\text{XC}}_{\alpha}$, 
we introduce two approximation to make equation~\ref{eq:corr_pi}
integrable.
First, we consider the adiabatic approximation, making it independent
of the frequency $\omega$. Second, we assume that it is only a weak
function of the coupling strength $\alpha$, so we can also neglect
this dependency. The exchange-correlation part can therefore be
expressed as
\begin{equation}
    f^{\text{XC}, \alpha}_{ia, jb}(\omega) \approx \alpha \frac{\partial^2 E}{\partial \rho_{ia} \partial \rho_{jb}} .
\end{equation}
This allows us to re-write equation~\ref{eq:pi_alpha}
as
\begin{equation}
\label{eq:adia}
    \boldsymbol{\Pi}_{\alpha}(\omega) = \boldsymbol{\Pi}_{0}(\omega) + 
    \alpha \boldsymbol{\Pi}_{0}(\omega) \left( \mathbf{V} + \mathbf{F}^{\text{XC}}  \right) \boldsymbol{\Pi}_{\alpha}(\omega).
\end{equation}
Using equation~\ref{eq:adia}, both integrations in equation~\ref{eq:corr_pi}
can be performed analytically, leading to 
\begin{equation}
\label{eq:finalcorr}
    E^{C} = \frac{1}{2} \left( \sum_n \Omega_n^{+} - \text{Tr}(\mathbf{A}') - \text{Tr}(\mathbf{A}'') \right), 
\end{equation}
where $\Omega_n^{+}$ represent all positive poles of the 
interacting response function. Note that strictly only
positive poles must be considered, as the full matrix is
traceless and time-reversal symmetric, i.e. all eigenvalues
are occurring in positive/negative pairs.
The eigenvalues $\Omega$ can be extracted from the 
eigenvalue problem
\begin{equation}
\label{eq:n6corr}
\begin{pmatrix}
\ebg{\phantom{-}\mathbf{A}'} &
\ebg{\phantom{-}\mathbf{B}'} &
\enbg{\phantom{-}\mathbf{C}^{T}} &
\enbg{\phantom{-}\mathbf{C}^{T}} \\

\ebg{-\mathbf{B}'} &
\ebg{-\mathbf{A}'} &
\enbg{-\mathbf{C}^{T}} &
\enbg{-\mathbf{C}^{T}} \\

\enbg{\phantom{-}\mathbf{C}} &
\enbg{\phantom{-}\mathbf{C}} &
\nbg{\phantom{-}\mathbf{A}''} &
\nbg{\phantom{-}\mathbf{B}''} \\

\enbg{-\mathbf{C}} &
\enbg{-\mathbf{C}} &
\nbg{-\mathbf{B}''} &
\nbg{-\mathbf{A}''}
\end{pmatrix}
\begin{pmatrix}
\ebg{X'} \\
\ebg{Y'} \\
\nbg{X''} \\
\nbg{Y''}
\end{pmatrix}
=
\Omega
\begin{pmatrix}
\ebg{X'} \\
\ebg{Y'} \\
\nbg{X''} \\
\nbg{Y''}
\end{pmatrix}
\text{.}
\end{equation}

Equation~\ref{eq:finalcorr} yields the total correlation energy
of the system. To extract electron-fermion correlation
energies, we evaluate equation~\ref{eq:finalcorr} 
also with the modified uncoupled matrix 
\begin{equation}
\label{eq:n6unc}
\begin{pmatrix}
\ebg{\phantom{-}\mathbf{A}'} &
\ebg{\phantom{-}\mathbf{B}'} &
\phantom{-}\mathbf{0} &
\phantom{-}\mathbf{0} \\

\ebg{-\mathbf{B}'} &
\ebg{-\mathbf{A}'} &
\phantom{-}\mathbf{0} &
\phantom{-}\mathbf{0} \\

\phantom{-}\mathbf{0} &
\phantom{-}\mathbf{0} &
\nbg{\phantom{-}\mathbf{A}''} &
\nbg{\phantom{-}\mathbf{B}''} \\

\phantom{-}\mathbf{0} &
\phantom{-}\mathbf{0} &
\nbg{-\mathbf{B}''} &
\nbg{-\mathbf{A}''}
\end{pmatrix}
\begin{pmatrix}
\ebg{X'} \\
\ebg{Y'} \\
\nbg{X''} \\
\nbg{Y''}
\end{pmatrix}
=
\Omega^{\text{unc.}}
\begin{pmatrix}
\ebg{X'} \\
\ebg{Y'} \\
\nbg{X''} \\
\nbg{Y''}
\end{pmatrix}
\text{,}
\end{equation}
and obtain the electron-$^{13}$C and electron-$^{19}$F
correlation energies as
\begin{align}
    E^{\text{C}}_{e-f} = & E^{\text{C}} - E^{\text{C}}_{\text{unc.}} \text{,} \\
    \label{eq:rpa_energy_final}
    E^{\text{C}}_{e-f} = & \frac{1}{2} \sum_n \left( \Omega_n^{+} - \Omega_{n, \text{unc.}}^{+} \right).
\end{align}

The correlation energy between different fermions can therefore
be described as half the difference between the excitation
energies of the fully coupled and uncoupled systems.
We have not yet introduced approximations beyond the
frequency- and coupling strength independency of the 
terms involved in the calculation of $\mathbf{A}, \mathbf{B}$, 
and $\mathbf{C}$. The simplest approximation, 
neglecting $f^{XC}$ entirely, is labeled as direct RPA (dRPA). 
While being the most commonly used RPA variant
in electronic structure theory, it suffers from
self-interaction errors (SIE). Especially for nuclear wavefunctions 
originating from highly charged fermions, the SIE
can become extensive. Including $f^{XC}$ will be referred to
as ``vertex'' correction,
\cite{Olsen.Patrick.ea:Beyond.2019}
and ideally removes the SIE.
We will therefore test the impact
of different approximations in the blue and red blocks
shown in equations~\ref{eq:n6corr} and \ref{eq:n6unc}.

\subsection{Connection to Second-order Perturbation Theory}
\label{subsec:connection}

There is an interesting connection of equation~\ref{eq:rpa_energy_final}
to the second-order M\o{}ller-Plesset (MP2) electron-fermion correlation 
energy. If $A'$ and $A''$ are chosen to be diagonal, i.e. $A'=(\epsilon_{a'} - \epsilon_{i'})$
and $A''=(\epsilon_{a''} - \epsilon_{i''})$, and $B' = B'' = 0$, then
\begin{equation}
    E^{\text{C,diag}}_{e-f} = \frac{1}{2} \sum_n \left( \Omega_n^{+} - \Omega_{n, \text{unc.}}^{+} \right) = \sum_{i'a'j''b''} \frac{(i'a'|j''b'')^2}{\epsilon_{a'} + \epsilon_{b''} - \epsilon_{i'} - \epsilon_{j''}} \text{.}
\end{equation}
This equivalence can be proven using the Laplace-transformed variant, which yields
\cite{Almlof:Elimination.1991,Haser.Almlof:Laplace.1992A,Haser:Mller-Plesset.1993}
\begin{equation}
\label{eq:mp2_laplace}
    E^{\text{C, MP2}}_{e-f} = \sum_{i'a'j''b''} (i'a'|j''b'')^2 \int_0^{\infty} e^{-t(\epsilon_{a'} + \epsilon_{b''} - \epsilon_{i'} - \epsilon_{j''})} dt
    \text{.}
\end{equation}
Now we introduce the resolution-of-the-identity (RI) approximation 
\begin{equation}
    (i'a'|j''b'') = B_{i'a'}^P B_{j''b''}^P
    \text{.}
\end{equation}
Equation~\ref{eq:mp2_laplace} can then be re-written as
\begin{equation}
    E^{\text{C,MP2}}_{e-f} = \int_0^{\infty} dt \sum_{i'a'j''b''} \sum_{PQ} B_{i'a'}^P B_{j''b''}^P  B_{i'a'}^Q B_{j''b''}^Q e^{-t(\epsilon_{a'} + \epsilon_{b''} - \epsilon_{i'} - \epsilon_{j''})}
    \text{,}
\end{equation}
introducing the intermediates
\begin{align}
    R_{PQ}'(t) = & \sum_{i' a'} B_{i'a'}^P B_{i'a'}^Q e^{-t(\epsilon_{a'} - \epsilon_{i'})} \text{,} \\
    S_{PQ}''(t) = & \sum_{j'' b''} B_{j''b''}^P B_{j''b''}^Q  e^{-t(\epsilon_{b''} - \epsilon_{j''})}
    \text{,}
\end{align}
the Laplace-transformed electron-fermion MP2 correlation energy can finally be obtained as 
\begin{equation}
     E^{\text{C,MP2}}_{e-f} = \int_0^{\infty} dt \sum_{PQ} R_{PQ}'(t) S_{PQ}''(t) \text{.}
\end{equation} 
Now using Parseval's theorem, we obtain the following result
\begin{equation}
    E^{\text{C,RPA-diag}}_{e-f} = \frac{1}{2\pi} \int_0^{\omega}  d\omega \sum_{PQ} \widetilde{R}_{PQ}'(\omega) \widetilde{S}_{PQ}''(\omega)
    \text{,}
\end{equation}
where $\widetilde{R}_{PQ}'$ is the continuous Fourier transform of ${R}_{PQ}'$.
This conveniently allows for a $\mathcal{O}(N^4)$ scaling calculation of the 
MP2 electron-fermion correlation energy.
This result is especially convenient for the future development of 
double hybrid multicomponent/NEO density functionals 
\cite{Hasecke.Mata:Multicomponent.2025} as it shows that the
calculation of electron-fermion MP2 correlation energies is well 
justified even for DFT references.

\subsection{GW Correlation Self-energies in the Random-Phase Approximation}
\label{subsec:gw}

The correlation part of the $GW$ self-energy can be obtained
as outlined in ref.~\citenum{Holzer.Franzke:Beyond.2024}
using contour deformation as
\begin{equation}
 \label{eq:sigma_comb}
 \Sigma_{q}^{\text{C}} (\omega^{\text{F}}) = R^{\text{C}} _{q}(\omega^{\text{F}}) + I^{\text{C}} _{q}(\omega^{\text{F}}) \text{.}
\end{equation}
The terms $I_q^{\text{C}}(\omega^{\text{F}})$ and 
$R_q^{\text{C}}(\omega^{\text{F}})$, 
with the two parts $I$ and $R$ representing the 
integration over the imaginary axis ($I$)
\begin{align}
\label{eq:ContourI}
I^{\text{C}} _{q'}(\omega^{\text{F}}) = 
& -\frac{1}{4\pi} \sum_{p'} \int_{-\infty}^{\infty} d\omega' ~
\frac{W_{p'q',p'q'}^{\text{C}}(i\omega')}{\omega^{\text{F}} - \epsilon_{p'}^{\text{F}} + \omega'} 
\end{align}
and the
parts arising from Cauchy's residue theorem ($R$)
\begin{align}
\label{eq:ContourR}
R^{\text{C}} _{q'}(\omega^{\text{F}}) = & \sum_{p'} f_{p'} 
\left\{ W_{p'q',p'q'}^{\text{C}}(\omega^{\text{F}} - \epsilon_{p'}^{\text{F}}) \right\} \text{.}
\end{align}
$f_{p'}$ is the contribution of the residue, equal to either 
$\pm1, \pm0.5$ or $0$.\cite{Holzer.Klopper:Ionized.2019}
$\omega^{\text{F}}$  is the frequency with respect to the Fermi 
level, and we note
that the Fermi level must be chosen separately for each fermion.
The screened exchange $\mathbf{W}^{\text{C}}$ is subsequently constructed 
from the solution vectors \{$X',Y',X'',Y''$\} of the 
response matrix shown in equation~\ref{eq:n6corr} as described
in ref.~\citenum{Setten.Weigend.ea:GW-Method.2013}. Different levels
of vertex corrections can be built into $\mathbf{A}',\mathbf{B}'$ 
and $\mathbf{A}'',\mathbf{B}''$ in a straightforward manner by 
simply modifying the matrix elements.
Quasiparticle energies are subsequently obtained separately for each
fermion from the underlying Kohn--Sham eigenvalues and the diagonal 
part of the self-energy as
\begin{align}
\label{eq:QP_eq}
 \epsilon_{q'}^{\text{QP}} (\omega) = & \epsilon_{q'}^{\text{QP}} + 
 Z_{q'} \langle q' | \Sigma^{\text{X}}_{q'} + \Sigma^{\text{C}}_{q'}(\omega) - V^{\text{KS}}_{q'} |  q' \rangle \text{,} \\ 
  \epsilon_{q''}^{\text{QP}} (\omega) = & \epsilon_{q''}^{\text{QP}} + 
 Z_{q''} \langle q'' | \Sigma^{\text{X}}_{q''} + \Sigma^{\text{C}}_{q''}(\omega) - V^{\text{KS}}_{q''} |  q'' \rangle \text{.}
\end{align}

The quasiparticle energy of the nucleus thus corresponds to the 
energy required to remove it from the quantum mechanical system.
For the carbon and fluorine nuclei investigated in this work, 
this causes the final state to be highly charged.

\section{Computational details}

We consider the systems {$^{13}$C}, {CH$_4$}, {CClH$_3$}, {CCl$_3$H}, 
{CCl$_4$}, {CCl$_3$F}, and {CH$_2$ClF} for carbon and  {F}, {F$^-$}, 
{F$_2$}, {HF}, {CCl$_3$F}, {CH$_2$ClF}, {AuF}, and {TlH} for fluorine.
Initially, the geometries were optimized using standard
Born--Oppenheimer DFT using the CHYF functional. 
\cite{Holzer.Franzke:General.2025}
In the following step, multicomponent KS solutions
\cite{PhysRevLett.101.153001}
were obtained by treating both the carbon and fluorine nuclei 
as quantum mechanical objects. The CHYF functional was used 
again due to its general fermion design.
\cite{Holzer.Franzke:General.2025}
From the set of converged electronic and carbon/fluorine
KS eigenstates, the RPA correlation energies, 
$GW$ self-energies, and quasiparticle energies were subsequently
calculated. Correlation energies from the second-order M{\o}ller-Plesset 
(MP2) perturbation theory were obtained at the multicomponent
HF reference states. Further, MP2 calculations at the 
CHYF reference were used to validate the numerical equivalence
to diagonal RPA outlined in section~\ref{subsec:connection}.
All investigations were done using the uncontracted
aug-cc-pV6Z for carbon and fluorine, and the def2-QZVP basis set 
for the remaining elements if applicable. For Au and Tl, the x2c-TZVPPall-2c
basis sets were used.\cite{Pollak.Weigend:Segmented.2017}.
When X2C was used, 
\cite{Kutzelnigg.Liu:Quasirelativistic.2005}
a finite nucleus model centered at the
center-of-mass of the nuclear density was employed.
\cite{Chandra.He:finite-nucleus.1994,Franzke.Weigend:NMR.2019}
Basis sets for the quantum nuclei have been
optimized for the respective atoms and CCl$_3$F molecule.
The latter was chosen because of its relevance as
standard in nuclear magnetic resonance experiments, 
and additionally featuring low symmetry. For this purpose, 
a local development version of Turbomole 
\cite{Balasubramani.Chen.ea:TURBOMOLE.2020,Franzke.Holzer.ea:TURBOMOLE.2023} 
was extended to include analytical gradients for arbitrarily charged fermions, 
enabling the construction of near-optimal basis sets even in these initial 
investigations. The resulting segmented contracted 
\mbox{hextuple-$\zeta$} \mbox{[11s8p4d3f2g1h] 
$\rightarrow$[6s5p4d3f2g1h]} nuclear basis sets are reported in 
the electronic supporting information.
Auxiliary basis sets were generated using the 
automatic procedure \cite{Lehtola:Straightforward.2021,Lehtola:Automatic.2023}
of ERKALE \cite{Lehtola.Hakala.ea:ERKALE-A.2012} from a unity of the
electronic and  protonic basis sets. To consistently cover all
fitting cases in the auxiliary basis set, the electronic uncontracted 
aug-cc-pV6Z basis set was combined with the respective nucdef-6Z basis set. 
A threshold of 10$^{-8}$ was chosen in the Cholesky decomposition 
while generating the auxiliary basis set, yielding 
up functions with angular momentum up to $l=10$.\cite{Lehtola:Automatic.2023}
The auxiliary basis sets are reported in the electronic supporting 
information.

\section{Results}
\label{sec:results}

\subsection{Electron-Nucleus Correlation Energies}
\label{sec:encorr}

\begin{table}[htbp!]
    \caption{Definition of RPA variants used in this work. Name of the
    RPA variant is given. Reference denotes the used reference state/calculation, 
    and electron (blue), nucleus (red), and elec.-nuc. (green) describe 
    the corresponding approximation used to evaluate 
    equations~\ref{eq:n6corr} and \ref{eq:n6unc}.}
    \begin{tabular}{llccc}
    \toprule
    name & reference & \ebg{electron} & \nbg{nucleus} & \enbg{elec.-nuc.}  \\
    \midrule
    dRPA & CHYF & \ebg{dRPA} & \nbg{dRPA} & \enbg{RPA} \\ 
    dRPA+$\Gamma$  & CHYF & \ebg{dRPA} & \nbg{V+F$^{\text{XC}}$} & \enbg{RPA} \\ 
    diag+$\Gamma$ & CHYF & \ebg{diag} & \nbg{V+F$^{\text{XC}}$} & \enbg{RPA} \\
    MP2  & HF & \ebg{-} & \nbg{-} & \enbg{MP2} \\
    \bottomrule
    \end{tabular}
    \label{tab:rpa_variants}
\end{table}

To assess the electron-nucleus correlation energy, we use three 
different RPA variants as listed in table~\ref{tab:rpa_variants}.
For electron-electron interactions, we either use dRPA 
(``dRPA'' and dRPA+$\Gamma$ variants)
or a diagonal approach (diag+$\Gamma$). 
For the nucleus-nucleus interactions part, we use either dRPA or the 
fully vertex corrected TD-DFT kernel ($V+F^{XC}$) as outlined in section~\ref{subsec:N6corr}. 
Using the fully vertex corrected RPA kernel for the electron-electron interaction 
is prohibitively expensive, and therefore not considered. The electron-nucleus 
interaction is described using the pure Coulomb kernel, which we simply mark 
as ``RPA'' kernel due to the lack of exchange interactions in this case. 
As MP2 can be exactly partitioned into electron-electron, nucleus-nucleus, 
and electron-nucleus parts, no further distinction is needed in this case.

\begin{table}[htbp!]
    \caption{Calculated electron-$^{13}$C correlation energies for {$^{13}$C}, {CH$_4$}, {CClH$_3$}, {CCl$_3$H}, {CCl$_4$}, {CCl$_3$F}, and {CH$_2$ClF} at the RPA@CHYF level of theory. The nucdef-6Z basis set was assigned to the carbon quantum nucleus, while the carbon electrons were computed using the uncorrelated aug-cc-pV6Z. For the electrons of the remaining elements, the def2-QZVP basis set was employed. The nucdef-6Z basis set was chosen as the auxiliary basis set for carbon, while the other atoms were assigned with the def2-QZVP basis set. All values are given in mHartree.}
    \begin{tabular}{l
    S[table-format=-1.3]
    S[table-format=-1.3]
    S[table-format=-1.3]
    S[table-format=-1.3]
    S[table-format=-1.3]
    S[table-format=-1.3]
    S[table-format=-1.3]
    S[table-format=-1.3]}
    \toprule
     & \text{$^{13}$C} & \text{CH$_4$} & \text{CClH$_3$} & \text{CCl$_3$H} & \text{CCl$_4$} & \text{CCl$_3$F} & \text{CH$_2$ClF} \\
    \midrule
    dRPA          & -4.719 & -4.832 & -4.609 & -4.658 & -4.543 & -4.685 & -4.654 \\
    dRPA+$\Gamma$  & -4.933 & -5.053 & -4.815 & -4.894 & -4.772 & -4.894 & -4.866 \\
    diag+$\Gamma$ & -5.116 & -5.243 & -5.036 & -5.046 & -4.988 & -5.044 & -5.031 \\
    MP2           & -5.103 & -5.155 & -4.920 & -4.920 & -5.153 & -5.151 & -5.153 \\
    \bottomrule
    \end{tabular}
    \label{tab:corr_C_DFT}
\end{table}

\begin{figure}[t!]
    \centering
    \includegraphics[width=0.99\linewidth]{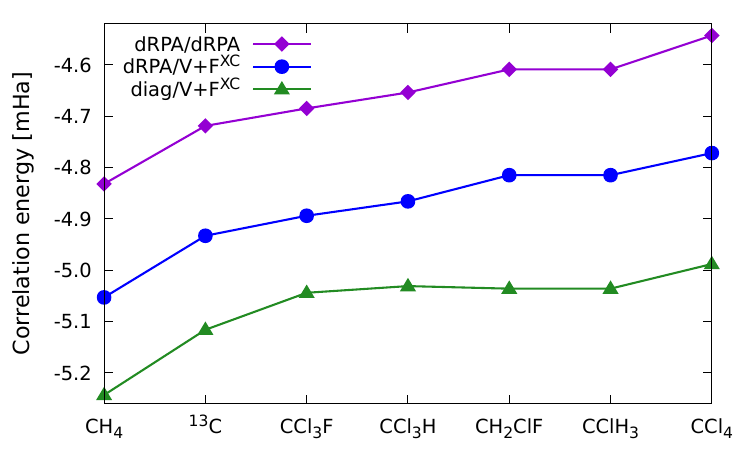}
    \caption{Calculated electron-$^{13}$C correlation energies for {$^{13}$C}, {CH$_4$}, 
    {CClH$_3$}, {CCl$_3$H}, {CCl$_4$}, {CCl$_3$F}, and {CH$_2$ClF} at the RPA@CHYF level of 
    theory for the dRPA (purple squares), dRPA+$\Gamma$  (blue circles), and diag+$\Gamma$ 
    (green triangles) approximations. All values in mHartree.}
    \label{fig:corr_C}
\end{figure}

Table~\ref{tab:corr_C_DFT} shows the electron-nucleus correlation energies obtained at the 
RPA@CHYF level of theory for the species containing $^{13}$C. Three different schemes were 
used to evaluate the correlation energies: The first line of results uses dRPA for both the 
electrons and the quantum nuclei. The second row uses dRPA for the electrons as well but 
contains the full vertex  contributions from the adiabatic exchange-correlation kernel for 
the quantum nuclei. The third line treats the quantum nuclei in the same manner, while diagonalizing 
the matrix for the electrons. Fig. \ref{fig:corr_C} compares the results of the three RPA 
methods visually. MP2 calculations were done as reference calculations to confirm the results. 
In general, the full dRPA yields correlation energies of the lowest magnitude,
which is in contrast to standard electronic dRPA results, which often overestimate correlation.
\cite{Eshuis.Bates.ea:Electron.2012,Holzer.Gui.ea:Bethe-Salpeter.2018}
This trend has been observed before for electron-proton correlation,\cite{Holzer.Franzke:Beyond.2024}
where the MP2 correlation energy is substantially larger than the dRPA one. 
This trend is also confirmed for the electron-$^{13}$C correlation, 
and table~\ref{tab:corr_C_DFT} outlines that indeed MP2 yields correlation energies 
larger by about 10~\% compared to dRPA. The dRPA results being too small is an artifact
of the large self-interaction error in the nuclear density part, leading to 
too large eigenvalues $\Omega^{\text{unc.}}$ in 
equation~\ref{eq:n6unc}. Subsequently, the Coulomb coupling in equation~\ref{eq:n6corr}
is overestimated, leading to lower than expected coupled eigenvalues $\Omega$, 
and incomplete error cancellation in the final electron-$^{13}$C correlation energy.
Correlation energies obtained from diag+$\Gamma$ for the quantum nuclei 
are the most negative ones, which is in line with the derivation shown in section~\ref{subsec:connection}, 
where it was outlined that MP2 and the diagonal approximation in the RPA are equivalent.
As the $^{13}$C nuclear density is highly localized, the vertex correction arising from 
$\mathbf{F^{\text{XC}}}$ is mainly canceling the Coulomb part. $\mathbf{F^{\text{XC}}}$ is 
dominated by the exact exchange contributions arising from the
CHYF functional. Inspecting the matrix elements of the $A''$ matrix indeed suggest the latter 
to be strongly diagonal dominant. The main difference between the diag+$\Gamma$@CHYF
and MP2@HF therefore arise from the different reference states. This has been confirmed by 
MP2@CHYF calculations, which yield basically identical results, with deviations of 
only 0.01-0.03~mHartree.
The dRPA+$\Gamma$ variant falls in between. It provides a more balanced
description between the correlation of the electrons and the fully
vertex-corrected response of the nuclear density. Most important, 
this variant mitigates the large self-interaction error obtained by 
a dRPA description of the nucleus, avoiding the over-cancellation
effect of the latter. At the same time, due to the coupling 
between the electronic and nuclear density parts, the influence of 
fluctuations in the electron density on the nuclear density 
are already recovered. We therefore deem this variant to be
the most accurate one.

For the $^{13}$C nucleus, the effect of the chemical surrounding 
is still significant, and can largely be explained by chemical 
intuition. CH$_4$ exhibits the largest (most negative) 
electron-$^{13}$C correlation energy, due to a slight increase in
electron density provided by the hydrogen atoms. The bare $^{13}$C atom sits between, 
and the addition of halogenides accordingly decreases the correlation energy
due to the slightly reduced electron density available at the nucleus.
However, these effects are small compared to those observed for hydrogen nuclei,
\cite{Holzer.Franzke:Beyond.2024} as the $1s$ core orbitals of $^{13}$C do not 
significantly contribute to bonding.
Contrary, chlorine is more electronegative than hydrogen, 
which reduces the electron density around the carbon nucleus in CCl$_4$ 
resulting in less electron-nucleus correlation.
The differences between various halogenation patterns are relatively subtle, 
however, asymmetric species such as CCl$_3$F and CH$_2$F$_2$ exhibit slightly 
larger correlation energies than the highly symmetric CCl$_4$. 
All tested RPA variants yield the same trend for 
$^{13}$C, as does MP2, as outlined in table~\ref{tab:corr_C_DFT}.

\begin{figure}[t!]
    \centering
    \includegraphics[width=0.99\linewidth]{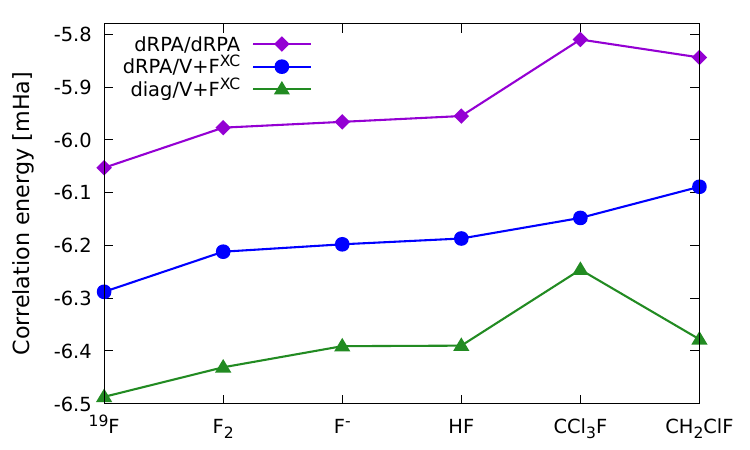}
    \caption{Calculated electron-$^{19}$F correlation energies for {F}, {F$^-$}, {F$_2$}, {HF}, {CCl$_3$F}, and {CH$_2$ClF} at the RPA@CHYF level of theory for the  dRPA (purple squares), dRPA+$\Gamma$  (blue circles), and diag+$\Gamma$ 
    (green triangles) approximations. All values in mHartree.}
    \label{fig:corr_F}
\end{figure}

For $^{19}$F, as outlined in table~\ref{tab:corr_F_DFT} and 
plotted in Fig. \ref{fig:corr_F}, the electron-$^{19}F$ correlation
energy gets larger in magnitude due to the more compact 1s orbital, 
induced by the higher charge. This compacting leads to an increase
in magnitude, but also to a flatter distribution between different 
molecular systems. Comparing the correlation energies of F and F$^-$, 
the F$^-$ electron-nucleus correlation energy has a lower magnitude, 
even though it features an additional electron. The increased diffuseness 
of the anion therefore leads to a decrease in the electron-$^{19}F$ correlation.
Notably, the pure dRPA as well as diag+$\Gamma$-RPA have problems in describing
the correlation for the CCl$_3$F molecule as outlined by figure~\ref{fig:corr_F}.
Only the dRPA+$\Gamma$ variant remedies this issue, removing the 
artificially low correlation energy in this case.
MP2@HF, on the other hand, yields results close to diag+$\Gamma$@CHYF, being
which is to be expected given the relation of these two methods
outlined in section~\ref{subsec:connection}. For the chlorinated
species CCl$_3$F and CH$_2$ClF however more pronounced differences 
can be seen, that we amount to the HF being unable to describe the
intrinsic correlation effects in the electron density.

\begin{table}[htbp!]
    \caption{Calculated electron-$^{19}$F correlation energies for {F}, {F$^-$}, {F$_2$}, {HF}, {CCl$_3$F}, and {CH$_2$ClF} at the RPA@CHYF level of theory. The nucdef-6Z basis set was assigned to the carbon quantum nucleus, while the carbon electrons were computed using the uncorrelated aug-cc-pV6Z. For the electrons of the remaining elements, the def2-QZVP basis set was employed. The F$_2$ value denotes the correlation energy per F nucleus. All values are given in mHartree.}
    \begin{tabular}{l
    S[table-format=-1.3]
    S[table-format=-1.3]
    S[table-format=-1.3]
    S[table-format=-1.3]
    S[table-format=-1.3]
    S[table-format=-1.3]}
    \toprule
    & \text{$^{19}$F} & \text{F$^-$} & \text{F$_2$} & \text{HF} &  \text{CCl$_3$F} & \text{CH$_2$ClF}  \\
    \midrule
    dRPA           & -6.053 & -5.966 & -5.977 & -5.955 & -5.810 & -5.844 \\ 
    dRPA+$\Gamma$   & -6.288 & -6.198 & -6.212 & -6.187 & -6.148 & -6.089 \\ 
    diag+$\Gamma$  & -6.487 & -6.391 & -6.431 & -6.390 & -6.246 & -6.378 \\
    MP2            & -6.456 & -6.443 & -6.446 & -6.348 & -6.444 & -6.443 \\
    \bottomrule
    \end{tabular}
    \label{tab:corr_F_DFT}
\end{table}

Finally, the influence of scalar relativistic effects on the electron-$^{19}F$ 
correlation has been calculated and is outlined in table~\ref{tab:corr_F_DFT_X2C}.
Scalar relativity leads to a contraction of the 1s orbitals, which in turn leads
to an increase in the magnitude of the electron-$^{19}$F correlation energy. 
The increase is in the order of 4-5~\%, but all further trends observed previously
are preserved. The bare F atom still has the largest magnitude, 
with the AuF and TlF fluorides exhibiting the expected 0.1~mHartree
decrease in correlation energy as previously observed also for HF.

\begin{table}[htbp!]
    \caption{Calculated electron-$^{19}$F correlation energies for 
    {F}, {AuF}, and {TlF} at the X2C-RPA@CHYF and MP2@HF level of theory. 
    The nucdef-6Z basis set was assigned to the fluorine quantum nucleus, 
    while the fluorine electrons were computed using the uncontracted 
    aug-cc-pV6Z basis set. For the electrons of the remaining elements, 
    the x2c-TZVPPall-2c basis set was employed. All values are given in mHartree.}
    \begin{tabular}{l
    S[table-format=-1.3]
    S[table-format=-1.3]
    S[table-format=-1.3]}
    \toprule
    &  \text{F} & \text{AuF} & \text{TlF} \\
    \midrule
    dRPA          & -6.300 & -6.198 & -6.207 \\ 
    dRPA+$\Gamma$  & -6.545 & -6.451 & -6.447 \\ 
    diag+$\Gamma$ & -6.757 & -6.662 & -6.673 \\
    MP2           & -6.724 & -6.686 & -6.686 \\
    \bottomrule
    \end{tabular}
    \label{tab:corr_F_DFT_X2C}
\end{table}

\subsection{Removing a Nucleus from Molecular Systems}

As initially outlined in ref.~\citenum{Holzer.Franzke:Beyond.2024} for hydrogen nuclei, 
$GW$ can be used to calculate de-protonization energies. Similarly, $GW$ can be used to calculate the
energy needed to remove the $^{13}$C and $^{19}$F nuclei from molecular systems.
Tables~\ref{tab:qp_C_DFT} and \ref{tab:qp_F_DFT} show the respective quasiparticle energies 
obtained at various levels of $GW$ theory. Specifically, we compare the one-shot 
$G_0W_0$ variant using dRPA response and dRPA+$\Gamma$ response, as well as the 
partially self-consistent ev$GW$ with dRPA+$\Gamma$ response.
It is immediately obvious that neglecting vertex correction leads to sizable
errors due to the large self-interaction error in the self-energy $\Sigma^{\text{C}}$,
which reach over 5000$\,$eV for both nuclei. Introducing vertex corrections reduced 
this by a few thousand eV - we explicitly stress that this is not a typo.

\begin{table*}[htbp!]
    \caption{Calculated quasiparticle energies with $Z=1$ for {$^{13}$C}, {CH$_4$}, {CClH$_3$}, {CCl$_3$H}, {CCl$_4$}, {CCl$_3$F}, and {CH$_2$ClF} at the $GW$@CHYF level of theory. The nucdef-6Z basis set was assigned to the carbon quantum nucleus, while the carbon electrons were computed using the uncorrelated aug-cc-pV6Z. For the electrons of the remaining elements, the def2-QZVP basis set was employed. All values are given in eV.}
    \begin{tabular}{ll
    S[table-format=-4.2]
    S[table-format=-4.2]
    S[table-format=-4.2]
    S[table-format=-4.2]
    S[table-format=-4.2]
    S[table-format=-4.2]
    S[table-format=-4.2]
    S[table-format=-4.2]}
    \toprule
     &  & \text{$^{13}$C}  & \text{CH$_4$} & \text{CClH$_3$} & \text{CCl$_3$H} & \text{CCl$_4$} & \text{CCl$_3$F} & \text{CH$_2$ClF} \\
    \midrule
     dRPA         & $\Sigma_c$ ($G_0W_0$) & 5293.54 & 5529.47 & 5358.11 & 5384.66 & 5426.76 & 5385.80 & 5352.20 \\
     dRPA+$\Gamma$ & $\Sigma_c$ ($G_0W_0 + \Gamma$) & 1279.65 & 1340.89 & 1356.10 & 1387.33 & 1402.36 & 1387.07 & 1352.79 \\
     dRPA+$\Gamma$ & $\Sigma_c$ (ev$GW +\Gamma$) & 1223.14 & 1314.14 & 1320.72 & 1343.29 & 1357.61 & 1341.97 & 1311.50 \\
    \midrule
     dRPA         & $\epsilon^{\text{QP}}$ ($G_0W_0$) & 2929.08 & 3148.04 & 2989.54 & 3038.65 & 3091.08 & 3054.22 & 2998.52 \\
     dRPA+$\Gamma$ & $\epsilon^{\text{QP}}$ ($G_0W_0 + \Gamma$) & -1084.81 & -1040.54 & -1012.47 & -958.68 & -933.33 & -944.51 & -1000.89 \\
     dRPA+$\Gamma$ & $\epsilon^{\text{QP}}$ (ev$GW +\Gamma$) & -1141.32 & -1067.29 & -1047.85 & -1002.72 & -978.08 & -989.60 & -1042.18 \\
    \bottomrule
    \end{tabular}
    \label{tab:qp_C_DFT}
\end{table*}

\begin{table*}[htbp!]
    \caption{Calculated quasiparticle energies with $Z=1$ for {F}, {F$^-$}, {F$_2$}, {HF}, {CCl$_3$F}, and {CH$_2$ClF} at the $GW$@CHYF level of theory. The nucdef-6Z basis set was assigned to the fluorine quantum nucleus, while the fluorine electrons were computed using the uncorrelated aug-cc-pV6Z. The x2c-TZVPPall basis set was assigned to Au and Tl. For the electrons of the remaining elements, the def2-QZVP basis set was employed. All values are given in eV.}
    \begin{tabular}{ll|
    S[table-format=-4.2]
    S[table-format=-4.2]
    S[table-format=-4.2]
    S[table-format=-4.2]
    S[table-format=-4.2]
    S[table-format=-4.2]
    S[table-format=-4.2]
    S[table-format=-4.2]}
    \toprule
      & & \text{F} & \text{F$^-$} & \text{F$_2$} & \text{HF} &  \text{CCl$_3$F} & \text{CH$_2$ClF} & \text{AuF (X2C)} & \text{TlF (X2C)} \\
    \midrule
    dRPA         & $\Sigma_c$ ($G_0W_0$) & 4991.72 & 5282.15 & 5106.05 & 5072.69 & 5219.11 & 5178.57 & 5316.31 & 5319.39 \\
    dRPA+$\Gamma$ & $\Sigma_c$ ($G_0W_0 + \Gamma$) & 3740.36 & 4069.51 & 3882.69 & 3862.04 & 4006.99 & 3965.89 & 4132.06 & 4134.97\\
    dRPA+$\Gamma$ & $\Sigma_c$ (ev$GW + \Gamma$) & 3605.06 & 4009.27 & 3721.68 & 3759.61 & 3893.57 & 3851.76 & 3994.98 & 4019.19 \\
    \midrule
    dRPA         & $\epsilon^{\text{QP}}$ ($G_0W_0$) & -1433.31 & -1363.77 & -1313.75 & -1372.56 & -1210.49 & -1261.63 & -1176.10 & -1189.54 \\
    dRPA+$\Gamma$ & $\epsilon^{\text{QP}}$ ($G_0W_0 + \Gamma$)  & -2684.67 & -2481.82 & -2537.22 & -2583.21 & -2422.61 & -2474.31 & -2360.35 & -2373.96 \\
    dRPA+$\Gamma$ & $\epsilon^{\text{QP}}$ (ev$GW + \Gamma$) & -2819.97 & -2542.06 & -2698.23 & -2685.64 & -2536.04 & -2588.45 & -2497.43 & -2489.74 \\
    \bottomrule
    \end{tabular}
    \label{tab:qp_F_DFT}
\end{table*}

Performing dRPA-based $GW$ calculations, which is the de-facto standard 
in electronic $GW$ calculations leads to catastrophic errors in multicomponent
calculations due to the significance of self-interaction errors.
A further discussion of these dRPA-based $G_0W_0$ results is therefore
neither needed nor meaningful within the context of this work. ev$GW$
is rendered impossible, as the large self-energy correction will lead to 
a breakdown of the Aufbau principle especially in the case of $^{13}$C
nuclei as outlined by the positive sign of $\epsilon^{\text{QP}}$ in
table~\ref{tab:qp_C_DFT}. The vertex correction outlined in section~\ref{subsec:N6corr} 
by design removes the self-interaction error, if the underlying functional
is correct for any one-particle density. The latter aspect is crucial, 
and also outlines that HF may be preferable to many density functional
approximations in this respect. HF, however, suffers from the absence of 
electron correlation, which we have already shown results in an incorrect 
ordering of correlation energies, particularly in the case of fluorine nuclei. 
This leaves one with a limited number of choices for reference densities
which can reasonably be used in a multicomponent $GW$ approach.
The dRPA+$\Gamma$-based $G_0W_0$ and ev$GW$ results bracket the 
reference quantum Monte Carlo energy of 2713.8$\,$eV,
\cite{Lüchow.Anderson:Accurate.1996} and the one obtained by 
summing all ionization energies (2715.9$\,$eV) for the fluorine atom, 
\cite{Biemont.Fremat.ea:IONIZATION.1999,NIST_ASD_5.12}
outlining that the vertex correction is indeed working as 
intended. Table~\ref{tab:qp_F_DFT} further shows that removing a 
fluorine nucleus becomes easier once it is part of a chemical system. 
This can be understood by noting that its removal leaves behind a highly charged system.
While removing the nucleus is energetically unfavorable in all investigated 
systems, the additional nuclei in molecular environments help stabilize 
the surplus electrons compared to the free atom. 
The highly charged Au and Tl nuclei
are especially ``effective'' in this respect, while a single hydrogen
atom in HF shows the least stabilizing effect.
For carbon, the chemical bonding situation is rather different. 
Carbon commonly occupies a central spot in the investigated molecules, 
and removing it from there is generally more favorable
if it is surrounded by electronegative bonding partner. 
Table~\ref{tab:qp_C_DFT}
even suggests that this effect is enhanced by chlorine
bonding partners, with $\epsilon^{\text{QP}}$ growing from
CH$_4$ to CClH$_3$ to CCl$_3$H to CCl$_4$.
Fluorine is better suited to stabilize additional
charge compared to chlorine. The results for the carbon atom itself should 
be treated with caution, as its electronic structure is relatively complex 
due to its multideterminant character, which likely leads to larger deviations 
than those observed in closed-shell compounds.
The sum of all ionization energies would amount to
1030.1~eV.\cite{NIST_ASD_5.12} $\epsilon^{\text{QP}}$
of the free $^{13}$C atom is significantly overestimating
this value. As a final remark, we note that Koopmans' theorem, 
which has been outlined as working reasonably well for hydrogen nuclei,
\cite{Schrader.Khanifaev.ea:Koopmans.2023}
is no longer compatible with heavier nuclei. The orbital energies of the HF and KS 
are again overestimated by more than 1000~eV, 
requiring significant quasiparticle corrections that are three 
orders of magnitude larger than observed for 
valence or core electrons.
\cite{Caruso.Dauth.ea:Benchmark.2016,Kehry.Klopper.ea:Robust.2023}
In light of these results, it is worth questioning whether 
applying Koopmans’ theorem to protons is truly justified.

\section{Conclusion}
\label{sec:conclusion}

We have derived a correlation energy expression generally valid 
for the interaction of electrons with fermionic nuclei. Being based on
the random-phase approximation, it yields a robust method to assess
these kind of interactions. Similarly to standard electronic RPA, 
the plasmonic formulas remain valid, and the correlation 
energy can be obtained as the difference of the interacting and 
non-interacting fermionic systems. Further, a unique equivalence of
minimally coupled RPA and second-order perturbation theory
for the interfermion correlation has been derived. 
The latter is only valid in this special context, but we assume it to 
be theoretically valuable for density functional
design, as it allows for the direct inclusion of perturbation theory
on general fermion (density) functionals.
Our numerical studies of the correlation between electrons and 
fermionic $^{13}$C and $^{19}$F nuclei reveal that the
correlation energy only slightly depends on the chemical environment, 
with the heavier $^{19}$F nucleus being less sensitive overall.  
Still, the higher nuclear charge of $^{19}$F leads to an enlarged
magnitude of the correlation energy overall, outlining that the 
denser distribution electrons outweighs the denser packing of the
nuclear density distribution. Scalar relativistic effects
on the correlation energy are found to again increase the magnitude of the 
correlation energy by 4-5~\% for the $^{19}$F nucleus due to the 
contraction of the electron density around the nucleus.
Numerical $GW$ simulations have proven to be possible, 
and we have shown that reasonable results can be obtained if
vertex corrections are used. Otherwise, self-interaction errors
become severe and will deter the results by up to a few thousand
eV, rendering them useless. Finally, the large self-energy corrections
outline that previous attempts of applying Koopman's
theorem to nuclear wavefunctions should be avoided.



\section*{Data Availability Statement}

The data that support the findings of this study are available within 
the article and its supplementary material.

\bibliography{library}

\end{document}